\documentclass{article}
\usepackage{spconf,amsmath,graphicx}
\usepackage[hidelinks]{hyperref}
\usepackage[utf8]{inputenc}
\usepackage[small]{caption}
\usepackage{amsthm}
\usepackage{booktabs}
\usepackage{algorithm}
\usepackage{algpseudocode}
\usepackage[switch]{lineno}
\usepackage{multirow}
\usepackage{makecell}
\usepackage{amsfonts}
\usepackage{subcaption}
\usepackage[table]{xcolor}
\usepackage{float}
\usepackage{stfloats}

\title{Leveraging Diverse Semantic-based Audio Pretrained \\Models for Singing Voice Conversion}
\name{
\begin{tabular}{c}
Xueyao Zhang$^1$\qquad
Zihao Fang$^1$\qquad
Yicheng Gu$^1$\qquad
Haopeng Chen$^1$ 
\\
Lexiao Zou$^2$\qquad
Junan Zhang$^1$\qquad
Liumeng Xue$^1$\qquad
Zhizheng Wu$^{1,\ddagger}$\thanks{$^{\ddagger}$Correspondence to \textit{wuzhizheng@cuhk.edu.cn}.}
\end{tabular}
}

\address{
$^1$The Chinese University of Hong Kong, Shenzhen\\
$^2$Shenzhen Research Institute of Big Data
}

\begin{document}
\ninept
\maketitle

\begin{abstract}
Singing Voice Conversion (SVC) is a technique that enables any singer to perform any song. To achieve this, it is essential to obtain speaker-agnostic representations from the source audio, which poses a significant challenge. A common solution involves utilizing a semantic-based audio pretrained model as a feature extractor. However, the degree to which the extracted features can meet the SVC requirements remains an open question. This includes their capability to accurately model melody and lyrics, the speaker-independency of their underlying acoustic information, and their robustness for in-the-wild acoustic environments. In this study, we investigate the knowledge within classical semantic-based pretrained models in much detail. We discover that the knowledge of different models is diverse and can be complementary for SVC.  Based on the above, we design a Singing Voice Conversion framework based on Diverse Semantic-based Feature Fusion (DSFF-SVC). Experimental results demonstrate that DSFF-SVC can be generalized and improve various existing SVC models, particularly in challenging real-world conversion tasks. Our demo website is available at \href{https://diversesemanticsvc.github.io/}{https://diversesemanticsvc.github.io/}.
\end{abstract}
\section{Introduction}
\label{sec:intro}
Singing Voice Conversion (SVC) aims to transform a singing signal into the voice of a target singer while maintaining the original lyrics and melody~\cite{svcc-paper}. This allows any singer to perform any song. It has a wide range of applications, such as music entertainment, singing voice enhancement, vocal education, and artistic creation.

\begin{figure}[t]
    \centering
    \begin{minipage}{\columnwidth}
         \includegraphics[width=\textwidth]{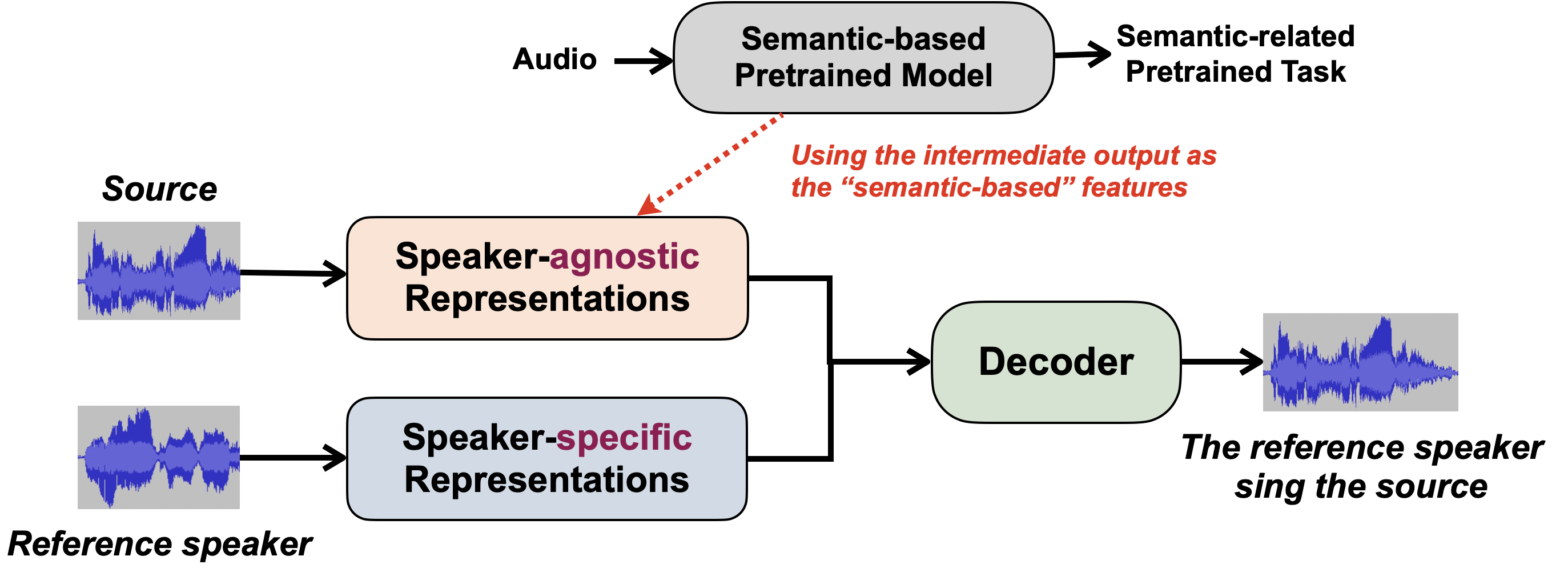}
         \caption{The role of semantic-based pretrained model in the classic singing voice conversion pipeline.}
         \label{fig:svc-pipeline}
    \end{minipage}
    \hfill
    \vspace{3mm}
    \begin{minipage}{\columnwidth}
        \centering
        \resizebox{0.9\textwidth}{!}{%
        \small
        \begin{tabular}{cccc}
        \toprule
        
        \makecell[c]{\textbf{Requirements of} \textbf{SVC}} & \multicolumn{3}{c}{\makecell[c]{\textbf{Capability of the} \textbf{Semantic-based Features}}}  \\
        
         \midrule
         To model melody & \multicolumn{3}{c}{\textit{Whether could or not} remains unknown} \\
         \midrule
         To model lyrics & \multicolumn{3}{c}{Could. But \textit{exactly how much} remains unknown} \\
         \midrule
         \makecell[c]{To model auxiliary \\acoustic information} & \multicolumn{3}{c}{\makecell[c]{Could. But \textit{whether the information is} \\ \textit{speaker-agnostic or not} remains unknown}} \\
         \midrule
         \makecell[c]{To be robust for in-the-wild\\ acoustic environment} & \multicolumn{3}{c}{\makecell[c]{\textit{Whether is robust or not} remains unknown}} \\ 
        \bottomrule
        \end{tabular}%
        }
        \captionof{table}{The extent to which the existing semantic-based features satisfy the requirements of singing voice conversion is still unclear.}
        \label{tab:requirements-svc}
    \end{minipage}
\end{figure}

In recent years, generative models~\cite{emilia,stable-diffusion} and conducting singing voice conversion with non-parallel data ~\cite{non-parallel-svc-facebook,non-parallel-svc-chenxin} has attracted more attention. 
Figure~\ref{fig:svc-pipeline} displays the classic pipeline for it. 
To empower the reference speaker to sing the source audio, the main idea is to extract the speaker-specific representations from the reference, extract the speaker-agnostic representations from the source, and then synthesize the converted audio using a decoder. Usually, speaker-specific representations can be just a one-hot speaker ID~\cite{authovc} or features extracted from a pretrained speaker verification model~\cite{authovc,zero-shot-svc-ismir}. For speaker-agnostic representations, common solutions involve using the intermediate output\footnote{The intermediate output is a dense high-dimensional vector instead of just symbolic token. It is believed that such output contains not only semantic but also acoustic information, which can enhance the quality of synthesized audio. In this paper, we refer to it as \textit{semantic-based} features.} from a semantic-based audio pretrained model. The pretraining task of this model is typically designed to be semantic-related, such as Automatic Speech Recognition (ASR)~\cite{non-parallel-svc-chenxin,ppg-based-svc} or self-supervised learning guided by semantics~\cite{self-supervised-vc,ssr-svc}.

For SVC, high-quality speaker-agnostic representations should meet several requirements (Table~\ref{tab:requirements-svc}). First, they should be capable of modeling melody and lyrics. Besides, they could also contain some auxiliary acoustic information (such as pronunciation, articulation, and prosody) to improve the naturalness and expressiveness of the synthesized audio. Last, they should be robust for varied acoustic environments of source audios (such as the in-the-wild singing voices separated from background music~\cite{zero-shot-roboust-svc-bgm,singfake}). However, for the semantic-based audio pretrained models, our understanding of their underlying knowledge remains limited, despite the considerable resources invested in pretraining~\cite{wav2vec2.0,hubert,whisper}. It is still unclear to what extent the extracted semantic-based features can satisfy the requirements of SVC. For example, except for the semantic signals, there is also acoustic information including prosody in the semantic-based features~\cite{acoustics-reference,tts-ssl-taslp}. Is this sufficient, or how much additional information is required to model the melody? Furthermore, is this information speaker-agnostic, or could it cause the source's timbre leakage~\cite{neucosvc} into the converted audio? Moreover, are these semantic-based pretrained models robust to diverse environments? How effective are these features when faced with in-the-wild audio data?

Motivated by this, by exploring the gap using only semantic-based features to conduct the SVC task, we investigate the underlying knowledge within the three classic semantic-based pretrained models respectively -- WeNet~\cite{wenet}, Whisper~\cite{whisper}, and ContentVec~\cite{contentvec}. Furthermore, we suppose that the capabilities of semantic-based features mainly depend on the pretraining tasks and data, and different pretraining ways will yield different underlying knowledge (Table~\ref{tab:three-content-features}). Based on this, we propose a concise and effective solution: by utilizing diverse semantic-based models that are pretrained distinctly, the capabilities of these jointly used semantic-based features will be enhanced for SVC. However, it is challenging to fuse the multiple features of diverse audio pretrained models. The reason is that the time resolutions of different models are usually mismatched, since their acoustic parameters like sampling rate and hop size can be different. To address it, we explored the efficacy and efficiency between a resampling strategy with cross attention~\cite{transformer,naturalspeech2}. The experiments verify that such signal processing based method could achieve the comparable performance to cross attention but with a lower latency and no additional training cost (Section~\ref{sec:resolution-align-results}). Based on the above, we design a Singing Voice Conversion framework based on Diverse Semantic-based Features Fusion (DSFF-SVC). Both objective and subjective evaluation results verify the effectiveness, generalization, and robustness of our proposed framework.


\begin{table*}[t]
\centering
\resizebox{\textwidth}{!}{%
\begin{tabular}{ccllllll}
\toprule
\textbf{School} & \textbf{Representative} & \textbf{Pretrained Task} & \textbf{Training Objective} & \textbf{Pretrained Data} & \textbf{Architecture}
\\ \midrule
\makecell[c]{\textbf{Supervised}} & \makecell[c]{{WeNet}\\ \cite{wenet}} & \makecell[l]{ASR, \textit{supervised} by \\linguistic labels} & \makecell[l]{CTC,\\Next Token Prediction \\ (character-level)} & \makecell[l]{Text-only transcription \\from Gigaspeech/Wenet-\\speech (10k hours English\\/Chinese speech)} & \makecell[l]{Encoder-decoder Conformer~\cite{conformer}\\ or Transformer~\cite{transformer}}
\\ \midrule
\makecell[c]{\textbf{Weak-Supervised}} & \makecell[c]{{Whisper}\\ \cite{whisper}} & \makecell[l]{Multitask including \\multilingual ASR, \\speech translation, \\spoken language identi-\\fication, etc., large-scale \\ \textit{weak} \textit{supervision}} & \makecell[l]{Next Token Prediction \\ (byte-level \cite{bpe})} & \makecell[l]{680k hours multilingual \\data, including both text-\\only and time-aligned\\ transcription} & \makecell[l]{Encoder-decoder Transformer~\cite{transformer}}
\\ \midrule
\makecell[c]{\textbf{Self-Supervised}} & \makecell[c]{{ContentVec}\\ \cite{contentvec}} & \makecell[l]{\textit{Self-supervised} learning \\which conditions on \\ disentangling speakers} & \makecell[l]{Masked Token \\Prediction (frame-level)} & \makecell[l]{Librispeech (1k hours \\English), using only speech\\ but no any transcription} & \makecell[l]{Encoder-only Transformer~\cite{transformer}}
\\ \bottomrule
\end{tabular}%
}
\caption{A systematic analysis for the three schools of the existing semantic-based pretrained models.}\label{tab:three-content-features}
\end{table*}

\section{Related Work}
\label{sec:Related Work}
The early singing voice conversion researches aim to design parametric statistical models such as HMM~\cite{parallel-svc-2009-HMM} or GMM~\cite{parallel-toda-2014,parallel-toda-2015} to learn the spectral features mapping of the parallel data. Since the parallel singing voice corpus is hard to collect on a large scale, the non-parallel SVC~\cite{non-parallel-svc-facebook,non-parallel-svc-chenxin}, or recognition-synthesis SVC~\cite{self-supervised-vc}, is popular in recent years. 
The classic pipeline of the non-parallel SVC is displayed in Figure~\ref{fig:svc-pipeline}.
To decouple speaker-agnostic and speaker-specific representations, some pioneering works leverage the information bottleneck~\cite{authovc} and the adversarial learning~\cite{non-parallel-svc-facebook,pitchnet} to design the end-to-end network. To introduce more explicit guidance instead of depending only on end-to-end learning, utilizing pretrained models as a feature extractor has become the trend. 

For the speaker-agnostic representations, the most well-known solution is to leverage the phonetic posteriorgrams (PPG) from pretrained ASR models as the semantic-based features~\cite{ppg-vc,ppg-based-svc}. In recent times, besides the supervised ASR models, an increasing number of semantic-based pretrained models have emerged under weak supervision and self-supervised learning. Researchers have explored in conducting the SVC task based on HuBERT~\cite{hubert,self-supervised-vc}, wav2vec 2.0~\cite{wav2vec2.0,ssr-svc}, Whisper~\cite{whisper,svcc-vits-ziqian} and more. 

However, the specific role of the semantic-based pretrained models and the extent to which the extracted features can meet the SVC requirements is still an open question. For example, some researchers point out that there is also rich acoustic information like prosody in these features~\cite{tts-ssl-taslp}, which is beneficial to improve the naturalness and the speaker similarity~\cite{acoustics-reference}. On the contrary, others believe that this acoustic information could be speaker-specific, which will cause the source's timbre leakage~\cite{neucosvc}. Motivated by this, we aim to investigate the specific knowledge within the semantic-based features by exploring to use them alone to conduct the SVC task. Besides, most existing works utilize a single source of semantic-based features. This study also researches whether different pretrained models can be complementary for SVC.

\section{Methodology}
\label{sec:method}
\begin{figure*}[hbt]
    \centering
    \includegraphics[width=\textwidth]{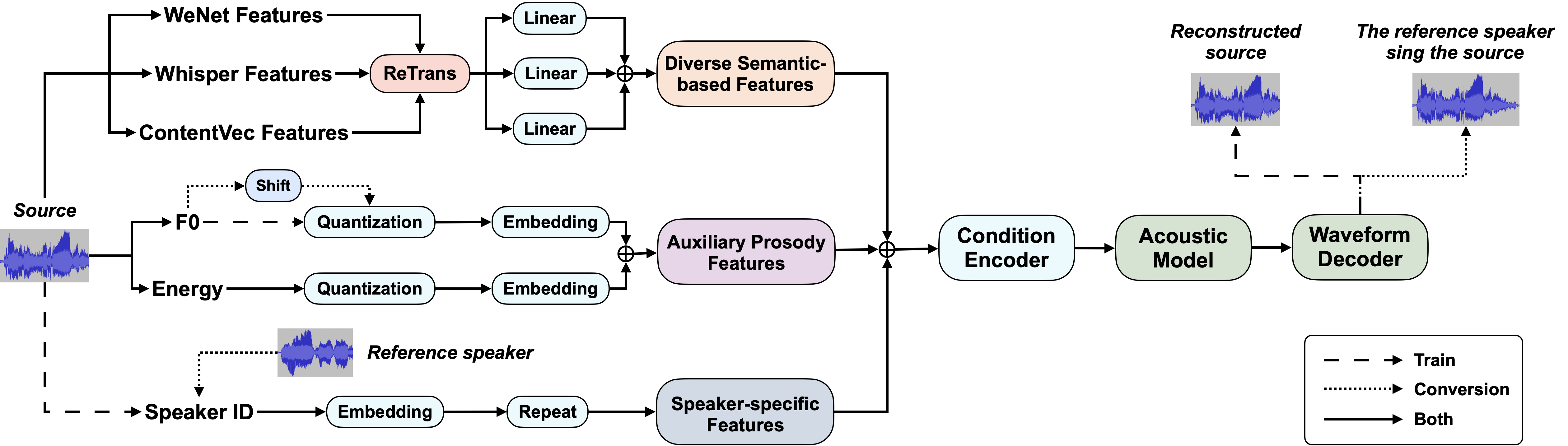}
    \caption{The proposed Singing Voice Conversion framework based on Diverse Semantic-based Features Fusion (DSFF-SVC). It is capable of incorporating most existing models (i.e., acoustic model and waveform decoder) as a base.}
    \label{fig:DSFF-SVC}
\end{figure*}

In this section, we analyze classic semantic-based pretrained models to demonstrate the potential knowledge embedded in semantic-based features. We also explain why combining multiple pretrained models can be effective for SVC (Section~\ref{sec:analysis-semantic-models}). Additionally, we discuss the use of resampling and cross-attention strategies for addressing the feature fusion issue of multiple pretrained models with different time resolutions (Section~\ref{sec:retrans}). Building on this, we introduce the Diverse Semantic-based Features Fusion SVC framework (DSFF-SVC, Figure~\ref{fig:DSFF-SVC}), which can integrate various models as its foundation.

\subsection{Analysis for Semantic-based Pretrained Models}
\label{sec:analysis-semantic-models}
In order for a model to capture the semantic signals of audio, we need to push its latent representations to align with semantic representations (such as characters or phonemes). In other words, the extent to which the latent representations contain semantic information depends on \textit{the level of semantic supervision} with which the audio model is pretrained. Furthermore, the knowledge of these latent representations that \textit{goes beyond semantic information} is also determined by the pretraining ways. Based on the above, we categorizes the existing semantic-based pretrained models into three schools (Table~\ref{tab:three-content-features}):
\begin{itemize}
    \item \textbf{Supervised models}: The models are pretrained under the ASR task, whose representative is WeNet~\cite{wenet}. For such models, the intermediate layers closer to the input (audio) contain more acoustic information, while layers closer to the output (character) contain more semantic information.
    \item \textbf{Weak-supervised models}: The models are pretrained under both ASR task and other auxiliary tasks, whose representative is Whisper~\cite{whisper}. Compared to supervised ones, such models are likely to contain additional knowledge related to the auxiliary tasks.
    \item \textbf{Self-supervised models}: The models conduct self-supervised learning under semantic (or speaker-agnostic) guidance, whose representative is ContentVec~\cite{contentvec}. The knowledge of these models is highly related to the construction of pseudo labels during the initial audio tokenization, since these labels usually play a role of ``teacher"~\cite{hubert}.
    Compared to the other two, the self-supervised ones contain more macro and general knowledge that is beneficial for modeling context.
\end{itemize}

In summary, the underlying knowledge of the three are likely to be different and diverse. In this study, we respectively select a representative for them -- WeNet~\cite{wenet}, Whisper~\cite{whisper}, and ContentVec~\cite{contentvec}, to explore the gap when adopting them into the SVC task. We will also investigate whether they can be complementary to each other.

\subsection{Features Fusion for Models of Mismatched Resolutions}
\label{sec:retrans}

When combining several pretrained audio models, a technical problem emerges because of the differing time resolutions of the models. In particular, for a specific utterance, the feature frame lengths extracted by different semantic-based pretrained models may vary. A classic signal processing based method for aligning features with different frame rates is to apply resampling. Besides of fusing the different semantic features, another common use case is to integrate a semantic-based model with a pretrained vocoder operating at different time resolutions.

\subsection{Singing Voice Conversion Framework based on Diverse Semantic-based Features Fusion}
\label{sec:DSFF-SVC}
The proposed Singing Voice Conversion framework based on Diverse Semantic-based Features Fusion (DSFF-SVC) is displayed in Figure~\ref{fig:DSFF-SVC}. The fusion of multiple features derived from diverse semantic-based pretrained models serves as the speaker-agnostic representation. A trainable embedding layer is employed to model the speaker features derived from a one-hot speaker ID. Furthermore, the incorporation of auxiliary prosody features, such as fundamental frequency (F0) and energy, facilitates the enhancement of melody modeling and the expressiveness of synthesized audio. 

For the semantic-based pretrained models, the WeNet, Whisper, and ContentVec are all Transformer-based~\cite{transformer} architecture. Here we utilize their encoder's output as the semantic-based features. Formally, given the utterance $u$ and the acoustic model $\mathcal{M}$, the extracted content features $\mathbf{c}_\mathcal{M} = \mathcal{M}(u) \in \mathbb{R}^{T_\mathcal{M} \times D_\mathcal{M}}$, where $\mathcal{M}$ can be either WeNet, Whisper, or ContentVec, $T_\mathcal{M}$ is the frame length, and $D_\mathcal{M}$ is the latent representation dimension of the model $\mathcal{M}$. Using resampling, we transform their frame lengths as the same (i.e., $\mathbf{\hat{c}}_{\mathcal{M}} \in \mathbb{R}^{T \times D_\mathcal{M}}$), and adopt adding fusion to merge them:

\begin{equation}
    \mathbf{\hat{c}} = \mathbf{Linear}(\mathbf{\hat{c}}_{\mathcal{M}_1}) \oplus \mathbf{Linear}(\mathbf{\hat{c}}_{\mathcal{M}_2}) \oplus \mathbf{Linear}(\mathbf{\hat{c}}_{\mathcal{M}_3})
\end{equation}
where $\mathbf{\hat{c}} \in \mathbb{R}^{T \times D}$, $\mathbf{Linear}$ means the linear layer, $\mathcal{M}_1$, $\mathcal{M}_2$, and $\mathcal{M}_3$ means WeNet, Whisper, and ContentVec, and $\oplus$ means the element-wise adding operation. 

For the auxiliary prosody features, we follow~\cite{diffsvc} to obtain the quantized F0 and energy features. We adopt the trainable embedding layers to get the F0 embeddings $\mathbf{f} \in \mathbb{R}^{T \times D}$ and energy embeddings $\mathbf{e} \in \mathbb{R}^{T \times D}$. For the speaker-specific representations, a look-up table with a trainable embedding layer is adopted to learn the speaker embeddings $\mathbf{s} \in \mathbb{R}^D$. 
Finally, we adopt the adding fusion to merge all the features:
\begin{table*}[t]
\centering
\resizebox{0.7\textwidth}{!}{%
\small
\begin{tabular}{lcccccccc}
\toprule
 \textbf{Semantic-based Features} & \textbf{MCD ($\downarrow$)} & \textbf{F0CORR ($\uparrow$)} & \textbf{F0RMSE ($\downarrow$)} & \textbf{CER ($\downarrow$)} & \textbf{SIM ($\uparrow$)}
\\ 
\midrule
Ground Truth & 0.000 & 1.000 & 0.0 & 12.9\% & 1.000 \\ \midrule
\rowcolor{gray!15}
 WeNet & 10.324 & 0.203 & 423.4 & 38.2\%  & 0.912 \\
 Whisper & \textit{8.229} & 0.524 & 297.3 & 18.9\%  & \textit{0.914} \\
 ContentVec & 8.972 & 0.491 & 361.0 & 22.1\% &  \textbf{0.918} \\
\midrule
\rowcolor{gray!15}
 WeNet + Whisper & 8.345 & 0.540 & 284.2 & \textit{16.8\%} & 0.911  \\
 WeNet + ContentVec & 8.870 & 0.525 & 329.5 & 19.9\% & 0.912 \\
 Whisper + ContentVec & \textbf{8.201} & \textit{0.548} & \textit{279.6} & 16.9\% & 0.912  \\
\midrule
\rowcolor{gray!15}
 WeNet + Whisper + ContentVec & 8.249 & \textbf{0.572} & \textbf{278.5} & \textbf{16.1\%} & 0.913  \\
\bottomrule

\end{tabular}%
}
\caption{Objective evaluation results of different semantic-based features and their integration. It can be observed that from WeNet to Wenet + Whisper, and then to Wenet + Whisper + ContentVec, the results of most metrics are promoted stage by stage.}\label{tab:results-content-features}
\end{table*}
\begin{equation}
    \mathbf{c} = \mathbf{CondEnc}(\mathbf{\hat{c}} \oplus \mathbf{f} \oplus \mathbf{e} \oplus \mathbf{\hat{s}})
\end{equation}
where $\mathbf{CondEnc}$ means a condition encoder that interacts with all the conditions of SVC. It is set as a simple linear layer in our experiments. $\mathbf{\hat{s}} \in \mathbb{R}^{T \times D}$ represents the frame-level speaker feature, being obtained by repeating $\mathbf{s}$ $T$ times.

During training, the DSFF-SVC conducts the reconstruction learning on the training corpus, and the speaker ID is just the speaker identity of every training sample. During conversion, given any source audio, we extract its semantic-based and energy features and stay them unchanged. To convert the speaker identity, we inject a reference speaker ID (which is seen in the training) to obtain the speaker-specific representations. For F0 features, we conduct the musical key transposition\footnote{\href{https://en.wikipedia.org/wiki/Transposition_(music)}{https://en.wikipedia.org/wiki/Transposition\_(music)}} to make the reference speaker sing the source song in his vocal range. Specifically, following~\cite{diffsvc,svcc-paper}, we shift the source F0 features by multiplying a factor, which is computed as the ratio between the F0 medians among the reference speaker's training corpus and the source audio.

Notably, the DSFF-SVC framework can support any architectures of acoustic models and waveform decoders (vocoders). During our experiments, we will investigate the generalization of DSFF-SVC for various base models including transformer-based, diffusion-based, and end-to-end models (Section~\ref{sec:expt-setup}).

\section{Experiments}
\label{sec:exp}

We conduct experiments to answer the following questions:
\vspace{5pt}
\begin{itemize}
    \item \textbf{EQ1}: How much the existing semantic-based pretrained models can meet the requirements of SVC? Could the multiple features from diverse semantic-based pretrained models be complementary?
    \item \textbf{EQ2}: How effective and generalized is the proposed singing voice conversion framework based on diverse semantic-based features fusion?
    \item \textbf{EQ3}: How effective is the resampling strategy compared to deep learning based method such as cross attention~\cite{transformer,naturalspeech2} for fusing multiple features of mismatched time resolutions?
\end{itemize}

\subsection{Experimental Setup}
\label{sec:expt-setup}
\vspace{5pt}
\subsubsection{Evaluation Tasks} 
We adopt two conversion settings to evaluate: (1) \textbf{Recording Studio Setting}: following the most existing works, we utilize the high-quality singing corpus that is recorded in studio as the experimental data. The vocals are clean with virtually no noise or environmental interference. Specifically, we use Opencpop~\cite{opencpop} as the target singer, whose training corpus is 5.2 hours of studio recorded singing voices. For source audios, we use M4Singer~\cite{m4singer} and randomly 25 utterances for each timbre type respectively (including \textit{Soprano}, \textit{Alto}, \textit{Tenor}, and \textit{Bass}).  (2) \textbf{In-the-Wild Setting}: in the real-world  SVC application, usually the singing voices are separated from the background music, which will remain some artifacts or reverb in the vocals~\cite{zero-shot-roboust-svc-bgm,singfake}. We consider this as a more challenging conversion task to examine the robustness of the SVC systems. Specifically, we adopt a private corpus which contains 6.4 hours of singing voices of 15 professional singers (6 English singers and 9 Chinese singers). The vocals are separated by Ultimate Vocal Remover (UVR)\footnote{\href{https://github.com/Anjok07/ultimatevocalremovergui}{https://github.com/Anjok07/ultimatevocalremovergui}}. We adopt four singers as the targets (an English male, an English female, a Chinese male, and a Chinese female). For source audios, we randomly sample 100 utterances which cover multiple musical genres including \textit{Pop}, \textit{Rock}, \textit{Folk} and \textit{Soul}.

\subsubsection{Evaluation Metrics}
For objective evaluation, following~\cite{svcc-paper}, we adopt Mel-cepstral distortion (MCD)~\cite{mcd}, F0 Pearson correlation coefficient (F0CORR), F0 Root Mean Square Error (F0RMSE), Character Error Rate (CER) which is obtained with the recognition results of whisper-large ASR model~\cite{whisper}. Besides, following~\cite{megatts}, we use the WavLM model~\cite{wavlm} finetuned for speaker verification\footnote{\href{https://huggingface.co/microsoft/wavlm-base-plus-sv}{https://huggingface.co/microsoft/wavlm-base-plus-sv}} to compute the cosine speaker similarity score (SIM).
For subjectivce evalution, we invite 12 volunteers who are experienced in the audio generation areas to conduct the Mean Opinion Score (MOS) evaluation in terms of naturalness and similarity. The naturalness score ranks from 1 (``Bad'') to 5 (``Excellent''), and the similarity score ranks from 1 (``Different speaker, sure'') to 4 (``Same speaker, sure'').

\subsubsection{Base Models} 
We select three base models to verify the generalization ability of DSFF-SVC framework: 
\vspace{5pt}
\begin{itemize}
    \item \textbf{TransformerSVC}: It adopts a vanilla encoder-only transformer model\cite{transformer} as the acoustic model. Its output is the mel-spectrogram.
    \item \textbf{VitsSVC}: It is a VITS-based~\cite{vits} model which is similar to the SoftVC-VITS\footnote{\href{https://github.com/svc-develop-team/so-vits-svc}{https://github.com/svc-develop-team/so-vits-svc}}. It is an end-to-end framework and can directly produce waveform.
    \item \textbf{DiffWaveNetSVC}: It adopts a diffusion-based acoustic model and could generate mel-spectrogram, which is proposed by~\cite{diffsvc}. The internal encoder of the diffusion framework is based on Bidirectional Non-Causal Dilated CNN\cite{diffwave}, which is similar to WaveNet\cite{Wavenet}.
\end{itemize}

\subsubsection{Implementation Details} 

For WeNet, we use the official models pretrained by 10k hours Wenetspeech\footnote{\href{https://github.com/wenet-e2e/wenet}{https://github.com/wenet-e2e/wenet}} to extract semantic-based features. For Whisper, we use the multilingual \textsc{MEDIUM} model\footnote{\href{https://github.com/openai/whisper}{https://github.com/openai/whisper}}. For ContentVec, we use the official {\textsc{500-CLASS} model} preatrained by 1k hours Librispeech\footnote{\href{https://github.com/auspicious3000/contentvec}{https://github.com/auspicious3000/contentvec}}. The latent space dimension $D$ is set as 384. We adopt Parselmouth~\cite{parselmouth} to extract F0 and compute L2-norm of the amplitude of each short-time Fourier transform frame as energy features~\cite{fastspeech2}. Following~\cite{diffsvc}, we set their numbers of bins for quantization as 256. For the three base models, we adopt the implementations of Amphion\footnote{\href{https://github.com/open-mmlab/Amphion/tree/main/egs/svc}{https://github.com/open-mmlab/Amphion/tree/main/egs/svc}}~\cite{amphion}. For DiffWaveNetSVC and TransformerSVC, we use the pretrained Amphion Singing BigVGAN\footnote{\href{https://huggingface.co/amphion/BigVGAN_singing_bigdata}{https://huggingface.co/amphion/BigVGAN\_singing\_bigdata}} as the vocoder to decode waveform from mel-spectrogram.
\vspace{5pt}
\subsection{Performance of Different Semantic-based Features (EQ1)}\label{sec:performance-different-content-features}
\vspace{3pt}
In this section, we aim to explore the gap using semantic-based features \textit{alone} to conduct the SVC task. We select DiffWaveNetSVC as the base model and conduct the task under the Recording Studio Setting. The experimental results are illustrated in Table~\ref{tab:results-content-features}. 

\begin{figure*}[t]
    \centering
    \begin{subfigure}[b]{0.28\textwidth}
         \centering
         \includegraphics[width=\textwidth]{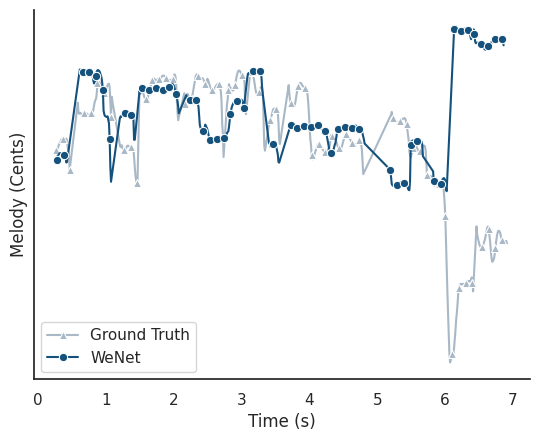}
         \caption{WeNet}
         \label{fig:f0-wenet}
     \end{subfigure}
     \begin{subfigure}[b]{0.28\textwidth}
         \centering
         \includegraphics[width=\textwidth]{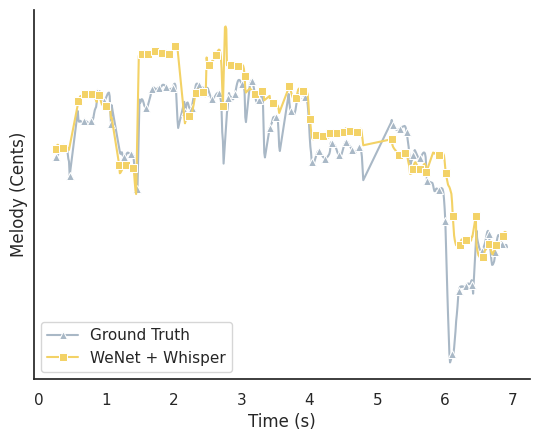}
         \caption{WeNet + Whisper}
         \label{fig:f0-wenet_whisper}
     \end{subfigure}
     \begin{subfigure}[b]{0.28\textwidth}
         \centering
         \includegraphics[width=\textwidth]{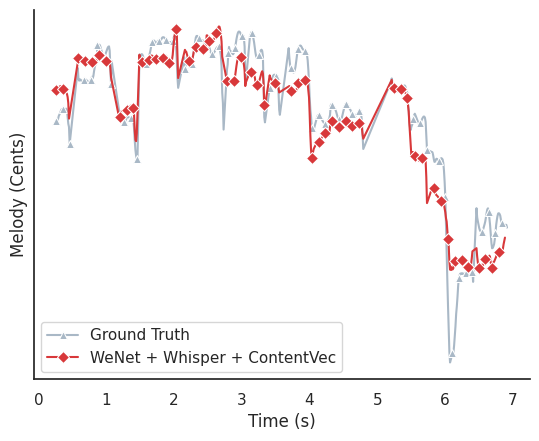}
         \caption{WeNet + Whisper + ContentVec}
         \label{fig:f0-wenet_whisper_contentvec}
     \end{subfigure}
     \caption{The complementary role of diverse semantic-based features in melody modeling. More benefits of the joint usage of diverse semantic-based features (including spectrogram reconstruction, lyrics modeling, etc.) can be seen at our \href{https://diversesemanticsvc.github.io/content_features.html}{\textit{\underline{demo website}}}.}
     \label{fig:melody-modeling-semantic-features}
\end{figure*}

On the one hand, the underlying knowledge of different semantic-based models can be distinct and diverse: (1) For modeling melody (F0CORR and F0RMSE), Whisper and ContentVec perform better than WeNet. This is mainly because the auxiliary tasks of the weak-supervised Whisper and the self-supervised ContentVec are beneficial for modeling more prosody-related signals. (2) For modeling lyrics (CER), Whisper performs best compared to the other two. It is assumed that the multilingual ASR tasks and the large-scale pretraining data allow Whisper to model more valid and robust semantic information. Moreover, ContentVec will provide better semantic signals than WeNet, although both are pretrained only on speech. It reveals that the self-supervised pretrained models could be more robust than the classical supervised ASR model. (3) When measuring speaker independence (SIM), the speaker similarity results of the three are comparable and all above 0.9, which is difficult to rank from human perception (see our \href{https://diversesemanticsvc.github.io/}{demo page}). This means that when using only semantic-based features for SVC, the three can all be considered speaker-agnostic.

On the other hand, the diverse semantic-based features can be complementary in most cases. For example, from WeNet to Wenet + Whisper, and then to Wenet + Whisper + ContentVec, the results of most metrics are promoted stage by stage. We display a case study of melody modeling in Figure~\ref{fig:melody-modeling-semantic-features}. It illustrates that after integrating diverse semantic-based features, the trajectories of the melody between converted audios and ground truth are closer. However, we can also find that using only semantic-based features is hard to model melody adequately (the highest F0CORR in Table~\ref{tab:results-content-features} is 0.572), appearing the ``out of tune" for human hearing. Therefore, introducing explicit melody modeling (such as F0 features) for SVC remains necessary in the present technology context.

\subsection{Performance of the DSFF-SVC framework (EQ2)}\label{sec:our-system}

\begin{table*}[htb]
    \centering
    \resizebox{\textwidth}{!}{%
    \begin{tabular}{clcccccccc}
    \toprule
    \multirow{2}{*}{\textbf{Base Model}} & \multirow{2}{*}{\textbf{Semantic-based Features}} & \multicolumn{4}{c}{\textbf{Recording Studio Setting}} & \multicolumn{4}{c}{\textbf{In-the-Wild Setting}} \\
    \cmidrule(lr){3-6} \cmidrule(lr){7-10}
    & & \makecell{\textbf{F0CORR} ($\uparrow$)} & \makecell{\textbf{F0RMSE} ($\downarrow$)} & \makecell{\textbf{CER} ($\downarrow$)} & \makecell{\textbf{SIM} ($\uparrow$)} & \makecell{\textbf{F0CORR} ($\uparrow$)} & \makecell{\textbf{F0RMSE} ($\downarrow$)} & \makecell{\textbf{CER} ($\downarrow$)} & \makecell{\textbf{SIM} ($\uparrow$)}  \\
    \midrule
    \multirow{3}{*}{\makecell[c]{\textbf{TransformerSVC}}} & WeNet & 0.849 & 149.3 & 15.6\% & 0.878 & 0.871 & 210.0 & 40.0\% & 0.865 \\
    & + Whisper & 0.924 & 77.2 & 14.9\% & 0.881 & 0.848 & 183.8 & 18.7\% & 0.867 \\
    & + Whisper + ContentVec & 0.931 & 75.5 & 16.2\% & 0.883 & 0.857 & 186.7 & 23.3\% & 0.868 \\
    \midrule
    \multirow{3}{*}{\textbf{VitsSVC}} & WeNet & 0.937 & 175.3 & 19.1\% & 0.890 & 0.919 & 91.3 & 57.7\% & 0.869 \\
    & + Whisper & 0.945 & 144.4 & 17.8\% & 0.890 & 0.920 & 86.9 & 35.2\% & 0.869 \\
    & + Whisper + ContentVec & 0.946 & 112.9 & 17.7\% & 0.886 & 0.921 & 79.5 & 32.3\% & 0.870 \\
    \midrule
    \multirow{3}{*}{\textbf{DiffWaveNetSVC}} & WeNet & 0.936 & 55.5 & 15.8\% & 0.875 & 0.901 & 87.8 & 60.8\% & 0.855 \\
    & + Whisper & 0.943 & 49.5 & 15.2\% & 0.884 & 0.921 & 73.6 & 21.1\% & 0.865 \\
    & + Whisper + ContentVec & 0.940 & 55.2 & 15.7\% & 0.884 & 0.919 & 79.9 & 23.3\% & 0.867 \\
    \bottomrule
    \end{tabular}%
    }
    \caption{Objective Evaluation Results of the proposed DSFF-SVC framework.}
    \label{tab:results-systems-obj}
\end{table*}

To verify the generalization and robustness of the proposed DSFF-SVC framework, we conduct experiments based on both the recording studio and the in-the-wild settings for the three base models. Here we also use the auxiliary prosody features (including F0 and energy), to improve the melody modeling and the expressiveness of the system. The objective and subjective evaluation results can be seen in Table~\ref{tab:results-systems-obj} and Table~\ref{tab:results-systems-sub}. It reveals that: (1) Conducting SVC under in-the-wild setting is more difficult. Facing to the more complex acoustic environment, only using the supervised WeNet produces a very poor performance. Integrating with the weak-supervised Whisper and the self-supervised ContentVec are usually beneficial, especially for Whisper. (2) For the generalization ability, we can observe that for all three models, the proposed idea of diversion semantic-based features fusion is effective in most cases under both settings. (3) For the robustness, particularly, under the in-the-wild setting, the improvement of integrating diverse semantic-based features is more obvious, especially for the objective CER and the subjective metrics.

Compared with Table~\ref{tab:results-content-features} and~\ref{tab:results-systems-obj}, there are also many interesting observations about auxiliary prosody features (F0 and energy): (1) After injecting such prosody signals into SVC models, the intelligibility (CER) has been improved a lot. In our opinion, this is because the prosody in singing voice is expressive and characteristic, which is hard to be modeled without the explicit assistant of prosody features like F0. Therefore, using only semantic-based features, some tone- and intonation-related signals will not be learned well, leading to the worse intelligibility. (2) Except for the intelligibility, it is also notable about the change of conversion speaker similarity (SIM). After introducinh F0 features, the melody-related metrics have been improved a lot and the converted singing voices are not out of tune any more. In the meantime, however, the speaker similarity is also harmed. We suppose such auxiliary prosody features own some speaker-specific signals, e.g. the personalized \textit{vibrato} patterns within F0 features. Therefore, directly copying the original prosody features from the source audio could not be the best choice for SVC, although it is a common way in the most SVC works. 

\begin{table}[H]
\centering
    \resizebox{\columnwidth}{!}{%
    \begin{tabular}{lcccc}
    \toprule
    \multirow{2}{*}{\makecell[l]{\textbf{Semantic-based} \textbf{Features}}} & \multicolumn{2}{c}{\textbf{Recording Studio}} & \multicolumn{2}{c}{\textbf{In-the-Wild}} \\
    \cmidrule(lr){2-3} \cmidrule(lr){4-5}
    & \textbf{Nat.} ($\uparrow$) & \textbf{Sim.} ($\uparrow$) & \textbf{Nat.} ($\uparrow$) & \textbf{Sim.} ($\uparrow$)  \\
     \midrule
    WeNet & 2.72 $\pm 0.22$ & 2.64 $\pm 0.21$ & 2.85 $\pm 0.21$ & 2.34 $\pm 0.20$ \\
    + Whisper & 4.02 $\pm 0.18$ & 3.13 $\pm 0.17$ & 3.70 $\pm 0.18$ & 2.86 $\pm 0.23$ \\
    \makecell[l]{+ Whisper + ContentVec} & 4.14 $\pm 0.19 $ & 3.25 $\pm 0.18$ & 3.71 $\pm 0.18$ & 2.82 $\pm 0.23$ \\
    \bottomrule
    \end{tabular}%
}
\caption{Subjective MOS evaluation results (with 95\% confidence interval) of DiffWaveNetSVC. The full scores of Naturalness (Nat.) and Similarity (Sim.) are 5 and 4.}\label{tab:results-systems-sub}
\end{table}
\subsection{Performance of the Resolution Transformation based Features Fusion (EQ3)}\label{sec:resolution-align-results}
\begin{table}[H]
\centering
\resizebox{\columnwidth}{!}{%
\small
\begin{tabular}{lcccccc}
\toprule
\multirow{2}{*}{\makecell[c]{\textbf{Fusion} \textbf{Strategy}}} & \multicolumn{2}{c}{\textbf{Computational Cost}} & \multicolumn{3}{c}{\textbf{Conversion Quality}} \\
\cmidrule(lr){2-3} \cmidrule(lr){4-6}
& \textbf{RTX} ($\uparrow$) & \textbf{RTF} ($\downarrow$) & \makecell{\textbf{F0CORR} ($\uparrow$)} & \makecell{\textbf{CER} ($\downarrow$)} & \makecell{\textbf{SIM} ($\uparrow$)}  \\
 \midrule
Cross attention & 191.9 & 2.57 & 0.896 & 16.4\% & 0.871  \\
Resampling & 415.0 & 0.86 & 0.936 & 15.8\% & 0.875 \\
\bottomrule
\end{tabular}%
}
\caption{The comparison between resampling and cross attention for features fusion.}\label{tab:results-resolution-align}
\end{table}
To verify the performance and efficiency of the resampling for feature fusing, we select the cross attention~\cite{transformer} as the compared method. Specifically, we utilize only WeNet to extract semantic-based features and try different strategies to align its resolution to that of F0 and energy features. We conduct the experiments on DiffWaveNetSVC and follow NaturalSpeech2~\cite{naturalspeech2} to adopt the cross attention for features fusion within the diffusion model.

Except for the quality evaluation for the conversion results, we also measure the computational cost of the both. Assume the duration of the whole training corpus is $d_{train}$, the duration of the evaluation samples to be converted is $d_{infer}$, the time to train an epoch is $t_{train}$, and the time to infer the evaluation samples is $t_{infer}$. We define RTX as $d_{train} / t_{train}$, which can measure the training efficiency. We define RTF as $t_{infer} / d_{infer}$ to measure the inference latency. The computational platform is a single NVIDIA Tesla V100. 

The comparison results between resampling and cross attention are displayed in Table~\ref{tab:results-resolution-align}. We can see that when fusing the features of mismatched time resolutions for the SVC task, resampling is even slightly better than cross attention in terms of conversion quality. Moreover, as a non-learning method, resampling does not introduce any additional parameters and greatly accelerates both training efficiency (RTX) and inference speed (RTF) much.

\section{Conclusion and Future Work}
This paper investigates three semantic-based pretrained models of supervised, weak-supervised, and self-supervised fashions for SVC. We discover their capabilities including modeling melody and lyrics are diverse and can be complementary. Based on these findings, we propose the SVC framework based on Diverse Semantic-based Features Fusion.  The experimental results confirm the effectiveness of our proposed framework, particularly for real-world applications involving in-the-wild data.
In addition, this paper raises other questions worth exploring.  Is it possible to pretrain one model that can capture both melody and lyrics signals while being speaker-agnostic at the same time? How should we select the pretraining method and construct the pretraining corpus to improve the robustness? We will leave these to our future research.

\bibliographystyle{IEEEbib}
\bibliography{strings,refs}

\begin{thebibliography}{10}

\bibitem{svcc-paper}
Wen-Chin Huang, Lester~Phillip Violeta, Songxiang Liu, Jiatong Shi, Yusuke Yasuda, and Tomoki Toda,
\newblock ``The singing voice conversion challenge 2023,''
\newblock in {\em {ASRU}}. 2023, {IEEE}.

\bibitem{emilia}
Haorui He, Zengqiang Shang, Chaoren Wang, Xuyuan Li, Yicheng Gu, Hua Hua, Liwei Liu, Chen Yang, Jiaqi Li, Peiyang Shi, Yuancheng Wang, Kai Chen, Pengyuan Zhang, and Zhizheng Wu,
\newblock ``Emilia: An extensive, multilingual, and diverse speech dataset for large-scale speech generation,''
\newblock in {\em Proc.~of SLT}, 2024.

\bibitem{stable-diffusion}
Robin Rombach, Andreas Blattmann, Dominik Lorenz, Patrick Esser, and Bj{\"{o}}rn Ommer,
\newblock ``High-resolution image synthesis with latent diffusion models,''
\newblock in {\em {CVPR}}. 2022, pp. 10674--10685, {IEEE}.

\bibitem{non-parallel-svc-facebook}
Eliya Nachmani and Lior Wolf,
\newblock ``Unsupervised singing voice conversion,''
\newblock in {\em {INTERSPEECH}}. 2019, pp. 2583--2587, {ISCA}.

\bibitem{non-parallel-svc-chenxin}
Xin Chen, Wei Chu, Jinxi Guo, and Ning Xu,
\newblock ``Singing voice conversion with non-parallel data,''
\newblock in {\em {MIPR}}. 2019, pp. 292--296, {IEEE}.

\bibitem{authovc}
Kaizhi Qian, Yang Zhang, Shiyu Chang, Xuesong Yang, and Mark Hasegawa{-}Johnson,
\newblock ``Autovc: Zero-shot voice style transfer with only autoencoder loss,''
\newblock in {\em {ICML}}. 2019, vol.~97, pp. 5210--5219, {PMLR}.

\bibitem{zero-shot-svc-ismir}
Shahan Nercessian,
\newblock ``Zero-shot singing voice conversion,''
\newblock in {\em {ISMIR}}, 2020, pp. 70--76.

\bibitem{ppg-based-svc}
Zhonghao Li, Benlai Tang, Xiang Yin, Yuan Wan, Ling Xu, Chen Shen, and Zejun Ma,
\newblock ``Ppg-based singing voice conversion with adversarial representation learning,''
\newblock in {\em {ICASSP}}. 2021, pp. 7073--7077, {IEEE}.

\bibitem{self-supervised-vc}
Wen{-}Chin Huang, Shu{-}Wen Yang, Tomoki Hayashi, and Tomoki Toda,
\newblock ``A comparative study of self-supervised speech representation based voice conversion,''
\newblock {\em {IEEE} J. Sel. Top. Signal Process.}, vol. 16, no. 6, pp. 1308--1318, 2022.

\bibitem{ssr-svc}
Tejas Jayashankar, Jilong Wu, Leda Sari, David Kant, Vimal Manohar, and Qing He,
\newblock ``Self-supervised representations for singing voice conversion,''
\newblock in {\em {ICASSP}}, 2023, pp. 1--5.

\bibitem{zero-shot-roboust-svc-bgm}
Naoya Takahashi, Mayank~Kumar Singh, and Yuki Mitsufuji,
\newblock ``Robust one-shot singing voice conversion,''
\newblock {\em arXiv}, vol. abs/2210.11096, 2022.

\bibitem{singfake}
Yongyi Zang, You Zhang, Mojtaba Heydari, and Zhiyao Duan,
\newblock ``Singfake: Singing voice deepfake detection,''
\newblock in {\em {ICASSP}}. 2024, {IEEE}.

\bibitem{wav2vec2.0}
Alexei Baevski, Yuhao Zhou, Abdelrahman Mohamed, and Michael Auli,
\newblock ``wav2vec 2.0: {A} framework for self-supervised learning of speech representations,''
\newblock in {\em NeurIPS}, 2020.

\bibitem{hubert}
Wei{-}Ning Hsu, Benjamin Bolte, Yao{-}Hung~Hubert Tsai, Kushal Lakhotia, Ruslan Salakhutdinov, and Abdelrahman Mohamed,
\newblock ``Hubert: Self-supervised speech representation learning by masked prediction of hidden units,''
\newblock {\em {IEEE} {ACM} Trans. Audio Speech Lang. Process.}, vol. 29, pp. 3451--3460, 2021.

\bibitem{whisper}
Alec Radford, Jong~Wook Kim, Tao Xu, Greg Brockman, Christine McLeavey, and Ilya Sutskever,
\newblock ``Robust speech recognition via large-scale weak supervision,''
\newblock in {\em {ICML}}. 2023, vol. 202, pp. 28492--28518, {PMLR}.

\bibitem{acoustics-reference}
Chao Wang, Zhonghao Li, Benlai Tang, Xiang Yin, Yuan Wan, Yibiao Yu, and Zejun Ma,
\newblock ``Towards high-fidelity singing voice conversion with acoustic reference and contrastive predictive coding,''
\newblock in {\em {INTERSPEECH}}. 2022, pp. 4287--4291, {ISCA}.

\bibitem{tts-ssl-taslp}
Chang Liu, Zhen{-}Hua Ling, and Ling{-}Hui Chen,
\newblock ``Pronunciation dictionary-free multilingual speech synthesis using learned phonetic representations,''
\newblock {\em {IEEE} {ACM} Trans. Audio Speech Lang. Process.}, vol. 31, pp. 3706--3716, 2023.

\bibitem{neucosvc}
Binzhu Sha, Xu~Li, Zhiyong Wu, Ying Shan, and Helen Meng,
\newblock ``Neural concatenative singing voice conversion: rethinking concatenation-based approach for one-shot singing voice conversion,''
\newblock in {\em {ICASSP}}. 2024, {IEEE}.

\bibitem{wenet}
Zhuoyuan Yao, Di~Wu, Xiong Wang, Binbin Zhang, Fan Yu, Chao Yang, Zhendong Peng, Xiaoyu Chen, Lei Xie, and Xin Lei,
\newblock ``Wenet: Production oriented streaming and non-streaming end-to-end speech recognition toolkit,''
\newblock in {\em INTERSPEECH}. 2021, pp. 4054--4058, {ISCA}.

\bibitem{contentvec}
Kaizhi Qian, Yang Zhang, Heting Gao, Junrui Ni, Cheng{-}I Lai, David~D. Cox, Mark Hasegawa{-}Johnson, and Shiyu Chang,
\newblock ``Contentvec: An improved self-supervised speech representation by disentangling speakers,''
\newblock in {\em {ICML}}. 2022, vol. 162, pp. 18003--18017, {PMLR}.

\bibitem{transformer}
Ashish Vaswani, Noam Shazeer, Niki Parmar, Jakob Uszkoreit, Llion Jones, Aidan~N. Gomez, Lukasz Kaiser, and Illia Polosukhin,
\newblock ``Attention is all you need,''
\newblock in {\em {NIPS}}, 2017, pp. 5998--6008.

\bibitem{naturalspeech2}
Kai Shen, Zeqian Ju, Xu~Tan, Yanqing Liu, Yichong Leng, Lei He, Tao Qin, Sheng Zhao, and Jiang Bian,
\newblock ``Naturalspeech 2: Latent diffusion models are natural and zero-shot speech and singing synthesizers,''
\newblock in {\em {ICLR}}. 2024, OpenReview.net.

\bibitem{conformer}
Anmol Gulati, James Qin, Chung{-}Cheng Chiu, Niki Parmar, Yu~Zhang, Jiahui Yu, Wei Han, Shibo Wang, Zhengdong Zhang, Yonghui Wu, and Ruoming Pang,
\newblock ``Conformer: Convolution-augmented transformer for speech recognition,''
\newblock in {\em {INTERSPEECH}}. 2020, pp. 5036--5040, {ISCA}.

\bibitem{bpe}
Rico Sennrich, Barry Haddow, and Alexandra Birch,
\newblock ``Neural machine translation of rare words with subword units,''
\newblock in {\em {ACL} {(1)}}. 2016, The Association for Computer Linguistics.

\bibitem{parallel-svc-2009-HMM}
Oytun T{\"{u}}rk, Osman B{\"{u}}y{\"{u}}k, Ali Haznedaroglu, and Levent~Mustafa Arslan,
\newblock ``Application of voice conversion for cross-language rap singing transformation,''
\newblock in {\em {ICASSP}}. 2009, pp. 3597--3600, {IEEE}.

\bibitem{parallel-toda-2014}
Kazuhiro Kobayashi, Tomoki Toda, Graham Neubig, Sakriani Sakti, and Satoshi Nakamura,
\newblock ``Statistical singing voice conversion with direct waveform modification based on the spectrum differential,''
\newblock in {\em {INTERSPEECH}}. 2014, pp. 2514--2518, {ISCA}.

\bibitem{parallel-toda-2015}
Kazuhiro Kobayashi, Tomoki Toda, Graham Neubig, Sakriani Sakti, and Satoshi Nakamura,
\newblock ``Statistical singing voice conversion based on direct waveform modification with global variance,''
\newblock in {\em {INTERSPEECH}}. 2015, pp. 2754--2758, {ISCA}.

\bibitem{pitchnet}
Chengqi Deng, Chengzhu Yu, Heng Lu, Chao Weng, and Dong Yu,
\newblock ``Pitchnet: Unsupervised singing voice conversion with pitch adversarial network,''
\newblock in {\em {ICASSP}}. 2020, pp. 7749--7753, {IEEE}.

\bibitem{ppg-vc}
Lifa Sun, Kun Li, Hao Wang, Shiyin Kang, and Helen~M. Meng,
\newblock ``Phonetic posteriorgrams for many-to-one voice conversion without parallel data training,''
\newblock in {\em {ICME}}. 2016, pp. 1--6, {IEEE} Computer Society.

\bibitem{svcc-vits-ziqian}
Ziqian Ning, Yuepeng Jiang, Zhichao Wang, Bin Zhang, and Lei Xie,
\newblock ``Vits-based singing voice conversion leveraging whisper and multi-scale f0 modeling,''
\newblock in {\em {ASRU}}. 2023, {IEEE}.

\bibitem{diffsvc}
Songxiang Liu, Yuewen Cao, Dan Su, and Helen Meng,
\newblock ``Diffsvc: {A} diffusion probabilistic model for singing voice conversion,''
\newblock in {\em {ASRU}}. 2021, pp. 741--748, {IEEE}.

\bibitem{opencpop}
Yu~Wang, Xinsheng Wang, Pengcheng Zhu, Jie Wu, Hanzhao Li, Heyang Xue, Yongmao Zhang, Lei Xie, and Mengxiao Bi,
\newblock ``Opencpop: {A} high-quality open source chinese popular song corpus for singing voice synthesis,''
\newblock in {\em {INTERSPEECH}}. 2022, pp. 4242--4246, {ISCA}.

\bibitem{m4singer}
Lichao Zhang, Ruiqi Li, Shoutong Wang, Liqun Deng, Jinglin Liu, Yi~Ren, Jinzheng He, Rongjie Huang, Jieming Zhu, Xiao Chen, and Zhou Zhao,
\newblock ``M4singer: {A} multi-style, multi-singer and musical score provided mandarin singing corpus,''
\newblock in {\em NeurIPS}, 2022.

\bibitem{mcd}
Robert Kubichek,
\newblock ``Mel-cepstral distance measure for objective speech quality assessment,''
\newblock in {\em Proceedings of IEEE pacific rim conference on communications computers and signal processing}. IEEE, 1993, vol.~1, pp. 125--128.

\bibitem{megatts}
Ziyue Jiang, Yi~Ren, Zhenhui Ye, Jinglin Liu, Chen Zhang, Qian Yang, Shengpeng Ji, Rongjie Huang, Chunfeng Wang, Xiang Yin, Zejun Ma, and Zhou Zhao,
\newblock ``Mega-tts: Zero-shot text-to-speech at scale with intrinsic inductive bias,''
\newblock {\em arXiv}, vol. abs/2306.03509, 2023.

\bibitem{wavlm}
Sanyuan Chen, Chengyi Wang, Zhengyang Chen, Yu~Wu, Shujie Liu, Zhuo Chen, Jinyu Li, Naoyuki Kanda, Takuya Yoshioka, Xiong Xiao, Jian Wu, Long Zhou, Shuo Ren, Yanmin Qian, Yao Qian, Jian Wu, Michael Zeng, Xiangzhan Yu, and Furu Wei,
\newblock ``Wavlm: Large-scale self-supervised pre-training for full stack speech processing,''
\newblock {\em {IEEE} J. Sel. Top. Signal Process.}, vol. 16, no. 6, pp. 1505--1518, 2022.

\bibitem{vits}
Jaehyeon Kim, Jungil Kong, and Juhee Son,
\newblock ``Conditional variational autoencoder with adversarial learning for end-to-end text-to-speech,''
\newblock in {\em {ICML}}. 2021, vol. 139, pp. 5530--5540, {PMLR}.

\bibitem{diffwave}
Zhifeng Kong, Wei Ping, Jiaji Huang, Kexin Zhao, and Bryan Catanzaro,
\newblock ``Diffwave: {A} versatile diffusion model for audio synthesis,''
\newblock in {\em {ICLR}}. 2021, OpenReview.net.

\bibitem{Wavenet}
A{\"{a}}ron van~den Oord, Sander Dieleman, Heiga Zen, Karen Simonyan, Oriol Vinyals, Alex Graves, Nal Kalchbrenner, Andrew~W. Senior, and Koray Kavukcuoglu,
\newblock ``Wavenet: {A} generative model for raw audio,''
\newblock in {\em {SSW}}. 2016, p. 125, {ISCA}.

\bibitem{parselmouth}
Yannick Jadoul, Bill Thompson, and Bart de~Boer,
\newblock ``Introducing {P}arselmouth: A {P}ython interface to {P}raat,''
\newblock {\em Journal of Phonetics}, vol. 71, pp. 1--15, 2018.

\bibitem{fastspeech2}
Yi~Ren, Chenxu Hu, Xu~Tan, Tao Qin, Sheng Zhao, Zhou Zhao, and Tie{-}Yan Liu,
\newblock ``Fastspeech 2: Fast and high-quality end-to-end text to speech,''
\newblock in {\em {ICLR}}. 2021, OpenReview.net.

\bibitem{amphion}
Xueyao Zhang, Liumeng Xue, Yuancheng Wang, Yicheng Gu, Xi~Chen, Zihao Fang, Haopeng Chen, Lexiao Zou, Chaoren Wang, Jun Han, Kai Chen, Haizhou Li, and Zhizheng Wu,
\newblock ``Amphion: An open-source audio, music and speech generation toolkit,''
\newblock {\em arXiv}, vol. abs/2312.09911, 2023.

\end{thebibliography}

\end{document}